# IMPACT OF THE ATMOSPHERIC BOUNDARY LAYER PROFILE ON THE VENTILATION OF A CUBIC BUILDING WITH TWO LARGE OPPOSITE OPENINGS

Alain BASTIDE, Franck LUCAS, Harry BOYER
Laboratoire de Génie Industriel Equipe Génie Civil Thermique de l'Habitat
40, Av de Soweto 97410 La Réunion - FRANCE
alain.bastide@univ-reunion.fr

## ABSTRACT

The aim of this paper is to show the influence of the atmospheric boundary layer profile on the distribution of velocity in a building having two large openings. The knowledge of the flow form inside a building is useful to define a thermal environment favourable with thermal comfort and good air quality. In computational fluid dynamics, several profiles of atmospheric boundary layer can be used like logarithmic profiles or power profiles. This paper shows the impact of these profiles on the indoor airflow. Non-ventilated or ventilated parts of room are found. They show respectively ineffective ventilation and effective ventilation. A qualitative and global approach allows to observe the flows in a cubic building and to show the influence of each profile according to the external ground roughness and the incidence angle of the wind. Some zones, where occupants move, are named volumes of life. Ventilation is there observed using traditional tools in order to analyze quantitatively the ventilation of these zones.

## INTRODUCTION

Under tropical countries, where climates are usually hot and humid, natural ventilation driven by winds is widely used in buildings. Generally, simplified methods using a modified Bernoulli equation are implemented in order to predict the aeraulic behaviour of a building. This method is based on the pressure coefficients ($C_P$) of external facade and on two other coefficients related to the airflow through each opening: the exponent of the flow (n) and the discharge coefficient ($C_D$). As an example Boyer et al. (Boyer et al, 1996, 1999), use it to predict the thermal and aeraulic behaviour of buildings in tropical countries. These models need some other parameters like air velocities to predict thermal comfort near occupants.

Under these latitudes, the means of electric energy production are limited and the use of active systems to reduce air temperature is misadvised. Some passive technical solutions are developed to create an acceptable situation of thermal comfort. For example, a classical technical solution is to open buildings. The first effect relating to the opening of the buildings is to increase the energy exchanges related to the buildings; the second effect is to increase the air velocity near occupants to improve the convective and evaporative exchanges. The aim of this last effect is to reach a situation of comfort or to stabilize an existing situation of comfort.

On the other hand, when one wishes to know the velocity distribution in building, it is necessary to implement tools using a finer space discretization like Computational Fluid Dynamics (CFD). Computational fluids mechanic allows, according to some approximations, to evaluate the pressure and velocity fields inside and outside buildings. Nevertheless, the consequence of this method is to generate a great number of information, which must be synthesized, with an aim of producing final information adapted to the posed problem of modelling. Techniques make it possible to reduce the complexity of the models, a priori or a posteriori, in order to obtain a simpler solution. These techniques break up into some categories. Find below, and following a non-exhaustive list, some techniques, a posteriori, of treatment used in the building literature:

- Statistical modeling;
- Classical modeling by means of standard deviation, mean, minimum and maximum;
- Qualitative interpretation of the numerical experimental.

Bastide (Bastide et al, 2004) treats the generation of adapted models based on the statistical modelling of the flows. In our case, our goal is to show that the flows vary according to the placed boundary conditions and not to develop a model. Nevertheless, a method presented in this last article is used in the continuation of the document. It, in particular, makes it possible to evaluate, in a quantitative way, the ventilation in the buildings.

Ernest (Ernest et al, 1991) treats modelling consisting in being interested only in mean velocities. Therefore, they study the variations of coefficients of velocity correlation ($C_V$) according to the ground roughness. Its experimentation takes place in a boundary layer wind tunnel. The air velocity profile at tunnel inlet boundary is logarithmic. They show that the variations of velocity coefficients are weak compared to the change of roughness.

Other studies, undertaken by: Kindangen (Kindangen et al, 1997) and Prianto (Prianto et al, 2003, 2002) use a qualitative way to express the advantages and disadvantages of architectural configurations.

Gouin (Gouin, 1984) and Ernest (Ernest et al,1991) study ventilation in a quantitative way (in a wind tunnel) while highlighting that ventilation depends on many parameters such as the facade porosity, the angle of incidence of the wind, the shape of the building. Ernest give a mathematical relation of the link, which exists between the field of pressure of facade, the porosity of facade, the angle of incidence and the turbulent phenomena.

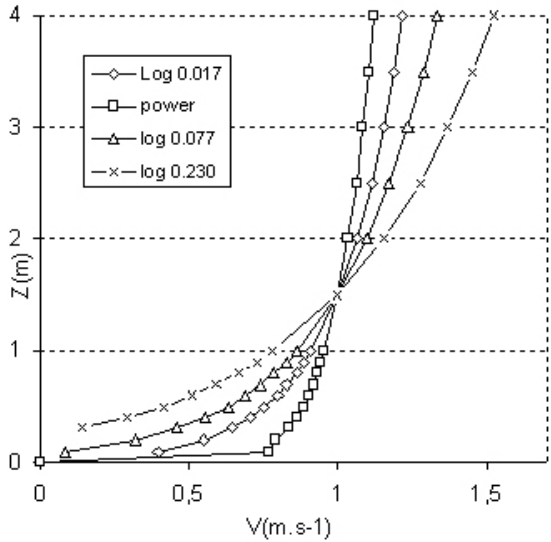

*Figure 1: Representation of various velocity profiles according to the ground roughness for the Log profile and a representation of a velocity power profile.*

On figure 1, various velocity profiles are represented with a same velocity reference: $U_{ref}(m.s^{-1})= 1 m.s^{-1}$ at 1.5m. The two velocity profiles used by Ernest et al. (Ernest et al, 1991) (related to two roughness lengths (z0): 0.017 and 0.077) are represented. We can note that these two profiles are rather similar in form. The profile in power corresponding to a length of roughness of 0.017 (according to Counihan, 1975) does not correspond to the logarithmic profile curve. They study weak variations of roughness of the external grounds and do not compare these results with experiments in hydraulic channel where the velocity profiles follow a power profile or do not compare results for significant variations of roughness.

It is clear that boundary layer profiles, based on ground roughness, strongly influence indoor air flow patterns. The aim of this paper is to quantify the effect or boundary layer profiles on such patterns through adapted mathematical tools.

## THEORETICAL BACKGROUND

**Atmospheric boundary layer - wind profiles**

The velocity profiles of wind depend on roughness of the grounds and a reference velocity: $V_{ref}$ measured with a height of $z_{ref}$. Various profiles are defined in the literature:

- logarithmic profile

$$U(z) = \frac{u_\tau}{\kappa} \log\left(\frac{z}{z_0}\right) \quad (1)$$

- power profile

$$U(z) = U_{ref}\left(\frac{z}{z_{ref}}\right)^{1/\alpha} \quad (2)$$

The parameter that formats the velocity profile is the roughness length ($z_0$) or the coefficient alpha ($1/\alpha$). The parameters which make possible to modify the reference velocity are the reference wind speed ($U_{ref}$) or the coefficient depending on the friction velocity ($u_\tau/\kappa$).

**Tools for ventilation evaluation**

At high Reynolds number the form of the flow is, on average, independent of the reference velocity of the atmospheric boundary layer. The mean velocity, in totality of the flow, is then directly proportional to the reference velocity. The quantitative way adopted in this paper use this specificity.

Consequently, it is possible to use the coefficient of velocity correlation $C_V$. This coefficient allowed, for example, in Gouin (Gouin, 1984), Kindangen (Kindangen, 1997), Ernest (Ernest, 1991), to evaluate architectural modifications on the indoor airflow. These authors calculated this coefficient at a fixed height, rather than particularizing its value for a specific, three-dimensional zone. Calculation of such a modified coefficient of velocity, as presented below, is made possible through an adapted CFD tool.

The expression of the coefficient is in the case of experimentation:

$$C_V = \sum_i \frac{U_i}{U_{ref}} \quad (3)$$

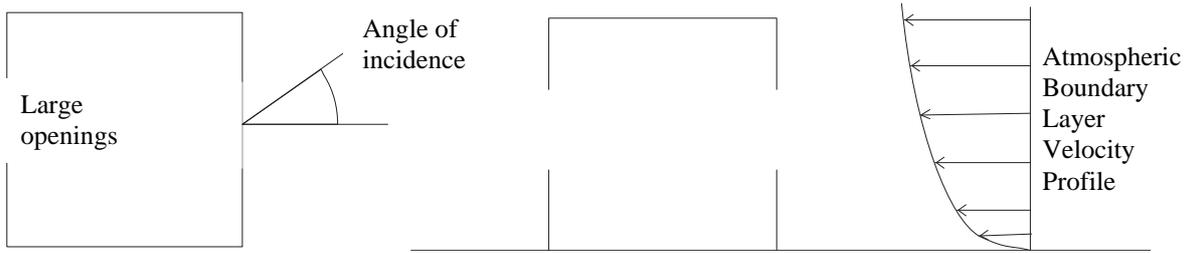

*Figure 2: Sketch of the CFD model*

This equation gives a specific average of values depending on the position of the sensors in the room.

The $C_V$ coefficient in the numerical case of experiments must be balanced by the cell volume of volume $V_i$ compared to total volume $V_t$. :

$$C_V = \sum_i \frac{U_i V_i}{U_{ref} V_t} \qquad (4)$$

This volume ($V_t$) is defined by the user to optimize ventilation there. i is the number of cells into the volume $V_t$.

**Volumes of life**

Aynsley (Aynsley, 1974) and the ASHRAE Handbook (ASHRAE, 2001) give air velocity ranges, within or beyond which comfortable or uncomfortable conditions are likely to occur. It is generally accepted that comfortable conditions can be maintained in the tropics with air velocities within the range of 0.3m/s to 1.0m/s; velocities beyond this last value tend to disturb loose paper in offices.

Thus to ensure comfortable situations in the tropics, it is essential to optimize air flow distributions within occupied environments. Traditionally, the indoor air velocity distribution is evaluated across a defined horizontal plane, as illustrated in Figure 3. This method does not provide however a means of characterizing the velocity distribution within a three-dimensional volume of an occupied room, e.g. around an individual. A better understanding of inter-zonal air flow patterns could provide guidance in window placement, office furniture settings, etc.

We refer to such a volume as volume of life, which can be characterized using a modified coefficient of velocity (Cv), correlated based on CFD calculations. This volume must be large enough to encompass the field of evolution of an individual.

Moreover, the volume of life must answer some criterions such as:

- The legs are not very sensitive to strong heats compared to the high part of the body.
- The volume of life must be sufficiently large to contain the field of evolution of the occupants.

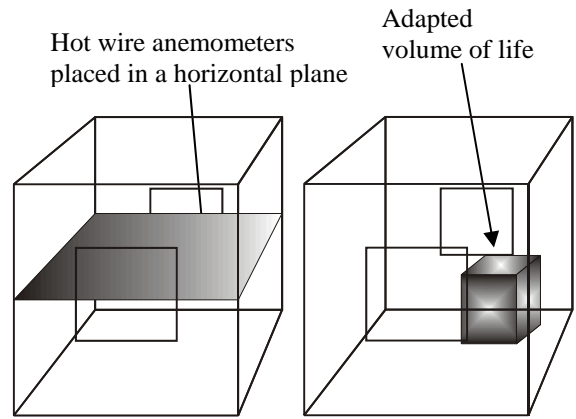

*Figure 3: Comparison of the two types of evaluation of the coefficients of velocity correlation: on the left: traditional technique and on the right: technique of volumes of life.*

## METHOD

**CFD model**

Several assumptions are required when using CFD. First, buoyancy effects are neglected. Second, surrounding terrain is considered unobstructed. Lastly, it is necessary to choose turbulence model adapted to the resolution of turbulent field. RNG-k-e is hence used.

The solution domain (15m x 15m x12m), including an empty room (3.0m x 3.0m x 3.0m), is shown in figure 2. The dimensions and positions of rectangular openings are fixed, only angle of incidence and ground roughness are changed. The flow and the cibic building are considered like isothermal. Modelling is steady-state in all simulations. Isothermal flow is assumed, with a fixed air temperature of 25°C. A rectangular, non-uniform mesh is chosen with small cells close to all surfaces and to the openings.

**Analysis of CFD results**

In this section, some CFD results are presented to show how the change in ground roughness has little effect on indoor airflow patterns.

The figure 4 shows the streamlines through the building. The cut is done along the symmetry axis passing by the center of the windows. The streamlines represented are thus calculated starting from the components U and W.

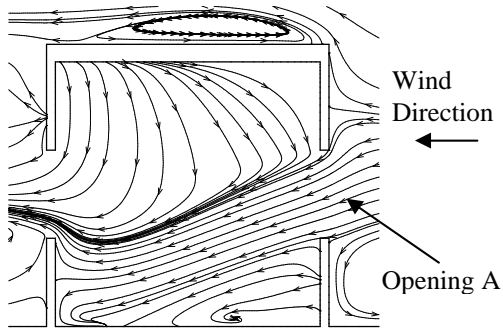

*Figure 4: Streamlines visualizing the airflow through the openings ($z_0=0.017$).*

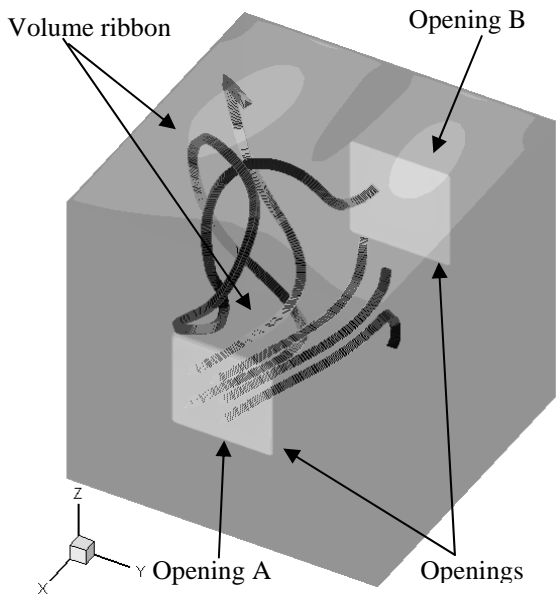

*Figure 5: Airflow pattern in the room.*

This figure 4 shows that, compared to the traditional results of air flow through such a building, the streamlines tend to curve towards the low part of the room. The traditional results show that the streamlines cross the part by the openings without deviating of their trajectory, whereas in this case we can notice which has a very different course. The difference results from the variation in external grounds roughness and from the velocity profile on the atmospheric boundary layer.

In Figure 5, the flow crosses the room by entering opening A and leaving through B. Ribbons are drawn to illustrate that certain streamlines do cross the room without much deviation, while others tend to follow more complex paths. The recirculation found near the ceiling is characterized by much lower velocities than in the bottom half of the room."

The flow tends to cross the room in the low part. The high part is, as for it, the place of great recirculations. These recirculations evolve at low speeds. Evaluated air velocities are much lower than in the low part.

## RESULTS

**Influence of atmospheric boundary layer profile on indoor streamlines**

Knowing that the velocity profiles of the atmospheric boundary layer vary according to the chosen type, we were interested in the comparison of the streamlines in the building for null incidence angle.

Initially, we can see that the streamlines are overall similar (Table 1). The type of profile of the atmospheric boundary layer does not condition in a fundamental way the form of the indoor streamlines. On the other hand, locally, we can note that the power profile induces a more significant slope to the bottom of the part. This can be notice in the bottom of the window B.

| logarithmic | power |
|---|---|
| $z_0=0.077$ | $1/\alpha=0.14$ |

*Table 1: Streamlines for the two profile types: on the left, power and on the right, logarithmic.*

Now, if we are interested in two profiles corresponding to really different roughnesses (plain $z_0=0.017$ and urban environment $z_0=0.154$, see Table 2), then we note also that the streamlines located between the openings follow practically the same way. The recirculation zones of the high part of the room are almost identical in both cases.

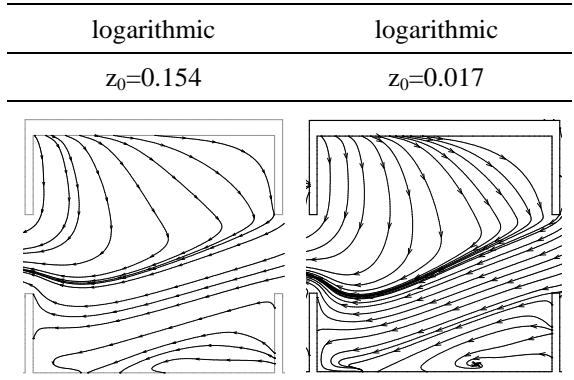

| logarithmic | logarithmic |
| --- | --- |
| $z_0=0.154$ | $z_0=0.017$ |

*Table 2: Streamlines for the two roughness length*

The tests carried out on this same building and for values of incidence angle ranging between 0° and 90° show they that the variations are weak.

**Correlation of Velocity Coefficient**

*Low part of the room*

Consider a person lying on a bed. As the bed can be place anywhere within the room and considering that surrounding furnishings have little influence on air flow, the indoor volume of interest (i.e. volume of evolution of the occupant) is defined between 0.2m and 0.6m in height, while horizontal limits are taken 0.2m from surrounding walls.

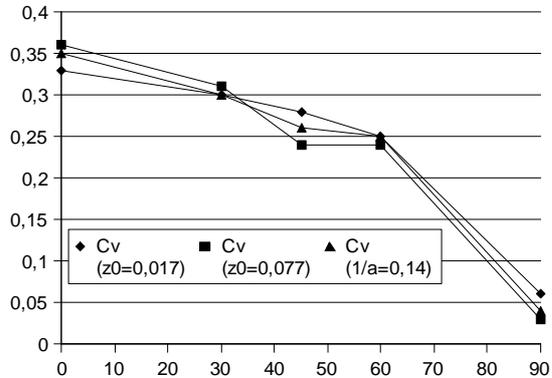

*Figure 6: Variations of coefficient $C_V$ according to the roughness of the external grounds.*

The general shape of the curves (Figure 6) is characteristic of the results generally obtained when coefficients $C_V$ is represented. Best ventilation is obtained for angles of incidence close to zero. As the angle of incidence increases, the value of the coefficient decreases and tends towards values close to zero.

In this volume of life, ventilation is to the maximum of about 30% the reference speed ($U_{ref}$). For the incidence angle of 90°, the values of $C_V$ are not null, whereas the wind does not enter directly the building. It's cause by turbulent phenomena and instabilities, which generate indoor air movement.

The variations of $C_V$ are weak (less than 10%) between each value for a given angle of incidence.

*A volume adapted to a worker*

We are interested now in a volume adapted to a person who works upright in this room. Volume is defined, according to the vertical, so that ventilation is optimal on the face level. It corresponds to the representation made on figure 3.

The results show that the coefficients of velocity correlation are not affected by the change of roughness of the external grounds. The variations observed do not exceed the 7% although the velocity profile evolves in each simulation.

*A volume adapted to a worker moving in all the part.*

Previous volume is shown. It is wide according to the horizontal one with all the room.

Variations in this last case do not excess 15% between each configuration of atmospheric boundary layer. On the other hand this greater variation, is explained by the variation of the flow close to the opening B.

Nevertheless, we can consider that the impact of the modifications of the roughness of the grounds is negligible in comparison with the results.

## CONCLUSION

At high Reynolds number, the form of the flow is independent of the intensity the velocity reference of the atmospherical boundary layer. Ernest (Ernest, 1991) shows that for close roughnesses, the bearing results on the coefficient of velocity correlation are similar.

As Ernest, for very different roughnesses or different modelings (profile log or prog power), we could highlight by this study that this parameter does not influence, in a notable way, the indoor flow.

During the analysis of flows inside opened buildings, it is always posed the question of the choice of the boundary condition and the repercussion of this choice about the solution. Using this study and for the building described here, we can conclude that the parameters of the boundary layer does not influence the non-dimensional coefficient $C_V$ and on the form of the flow.

On the other hand, the flow in this building slightly seems to be inclined towards the low part of the room, of this fact, for buildings deeper (where the distance between the openings is larger) the impact of the shape of the boundary layer is likely to be an influential parameter. By consequence, the non-dimensionnal coefficient $C_V$ risk to be affected.

Moreover, the profiles of atmospheric boundary layers used are ideal profiles where the grounds are regarded as smooth and the profiles are defined on grounds without asperities. However, it is clear, that during brutal modifications of roughness the profile of boundary layer evolves according to the vertical. Then the interior flow is, it also, modified.

A thorough study is under development to show than the other flow parameters, related in particular to the flows ($\dot{m}$, $C_D$) do not vary significantly.